\documentclass{isprs} 
\usepackage[utf8]{inputenc}

\usepackage{subfigure}
\usepackage{setspace}
\usepackage{geometry} 
\usepackage{epstopdf}
\usepackage[labelsep=period]{caption}  
\usepackage[british]{babel} 

\usepackage{amsmath,graphicx}
\usepackage{color}
\usepackage{tikz}
\usepackage{balance}
\usepackage{url}
\usepackage{multirow,hhline}
\usepackage{placeins}
\usepackage{wasysym}
\usepackage{amssymb}

\usepackage[nomessages]{fp}
\usepackage{array}
\renewcommand{\aa}[1]{{\bf \textcolor{blue}{#1}}}

\newcommand{\labsty}[1]{{\footnotesize \textbf{#1}}}
\newcommand{\labRef}{\labsty{Reference}}
\newcommand{\labNoise}{\labsty{Noisy}}
\newcommand{\labProp}{\labsty{Proposed}}
\newcommand{\labKldnn}{\labsty{KL-DNN}}
\newcommand{\labFans}{\labsty{FANS}}
\newcommand{\labSarbm}{\labsty{SAR-BM3D}}

\newcommand{\de}[2]{ \frac{\partial{#1}} {\partial{#2}} }
\newcommand{\image}{\pgfuseimage}

\newcommand{\mapspath}{./maps/}
\newcommand{\mercpath}{./mercedes/}
\newcommand{\rosenpath}{./Rosen/}

\newcommand{\figpath}{./network/}

\definecolor{mybegie}{RGB}{128,0,0}

\pgfdeclareimage[width=0.105\columnwidth]{sim1}{\mapspath /details/2_noisy.png}
\pgfdeclareimage[width=0.105\columnwidth]{ref1}{\mapspath /details/2_ref.png}
\pgfdeclareimage[width=0.105\columnwidth]{prop1}{\mapspath /details/2_prop.png}
\pgfdeclareimage[width=0.105\columnwidth]{kl1}{\mapspath /details/2_kldnn.png}
\pgfdeclareimage[width=0.105\columnwidth]{fans1}{\mapspath /details/2_fans.png}
\pgfdeclareimage[width=0.105\columnwidth]{sarbm3d1}{\mapspath /details/2_sarbm3d.png}

\pgfdeclareimage[width=0.105\columnwidth]{sim2}{\mapspath /details/5_noisy.png}
\pgfdeclareimage[width=0.105\columnwidth]{ref2}{\mapspath /details/5_ref.png}
\pgfdeclareimage[width=0.105\columnwidth]{prop2}{\mapspath /details/5_prop.png}
\pgfdeclareimage[width=0.105\columnwidth]{kl2}{\mapspath /details/5_kldnn.png}
\pgfdeclareimage[width=0.105\columnwidth]{fans2}{\mapspath /details/5_fans.png}
\pgfdeclareimage[width=0.105\columnwidth]{sarbm3d2}{\mapspath /details/5_sarbm3d.png}

\pgfdeclareimage[width=0.105\columnwidth]{sim3}{\mercpath /details/5_noisy.png}
\pgfdeclareimage[width=0.105\columnwidth]{ref3}{\mercpath /details/5_ref.png}
\pgfdeclareimage[width=0.105\columnwidth]{prop3}{\mercpath /details/5_prop.png}
\pgfdeclareimage[width=0.105\columnwidth]{kl3}{\mercpath /details/5_kldnn.png}
\pgfdeclareimage[width=0.105\columnwidth]{fans3}{\mercpath /details/5_fans.png}
\pgfdeclareimage[width=0.105\columnwidth]{sarbm3d3}{\mercpath /details/5_sarbm3d.png}

\pgfdeclareimage[width=0.15\textwidth]{ros_noisy}{\rosenpath //Rosen_1_noisy.png}
\pgfdeclareimage[width=0.15\textwidth]{ros_kldnn}{\rosenpath //Rosen_1_kldnn.png}
\pgfdeclareimage[width=0.15\textwidth]{ros_prop}{\rosenpath //Rosen_1_prop.png}
\pgfdeclareimage[width=0.15\textwidth]{ros_fans}{\rosenpath //Rosen_1_fans.png}
\pgfdeclareimage[width=0.15\textwidth]{ros_sarbm3d}{\rosenpath //Rosen_1_sarbm3d.png}

\pgfdeclareimage[width=0.15\textwidth]{ros_noisy1}{\rosenpath /details/1_noisy.png}
\pgfdeclareimage[width=0.15\textwidth]{ros_prop1}{\rosenpath /details/1_prop.png}
\pgfdeclareimage[width=0.15\textwidth]{ros_kldnn1}{\rosenpath /details/1_kldnn.png}
\pgfdeclareimage[width=0.15\textwidth]{ros_fans1}{\rosenpath /details/1_fans.png}
\pgfdeclareimage[width=0.15\textwidth]{ros_sarbm3d1}{\rosenpath /details/1_sarbm3d.png}

\pgfdeclareimage[width=0.15\textwidth]{ros_noisy2}{\rosenpath /details/1_noisy.png}
\pgfdeclareimage[width=0.15\textwidth]{ros_prop2}{\rosenpath /details/1_prop.png}
\pgfdeclareimage[width=0.15\textwidth]{ros_kldnn2}{\rosenpath /details/1_kldnn.png}
\pgfdeclareimage[width=0.15\textwidth]{ros_fans2}{\rosenpath /details/1_fans.png}
\pgfdeclareimage[width=0.15\textwidth]{ros_sarbm3d2}{\rosenpath /details/1_sarbm3d.png}

\pgfdeclareimage[width=0.15\textwidth]{ros_noisy2}{\rosenpath /details/2_noisy.png}
\pgfdeclareimage[width=0.15\textwidth]{ros_prop2}{\rosenpath /details/2_prop.png}
\pgfdeclareimage[width=0.15\textwidth]{ros_kldnn2}{\rosenpath /details/2_kldnn.png}
\pgfdeclareimage[width=0.15\textwidth]{ros_fans2}{\rosenpath /details/2_fans.png}
\pgfdeclareimage[width=0.15\textwidth]{ros_sarbm3d2}{\rosenpath /details/2_sarbm3d.png}

\pgfdeclareimage[width=0.65\columnwidth]{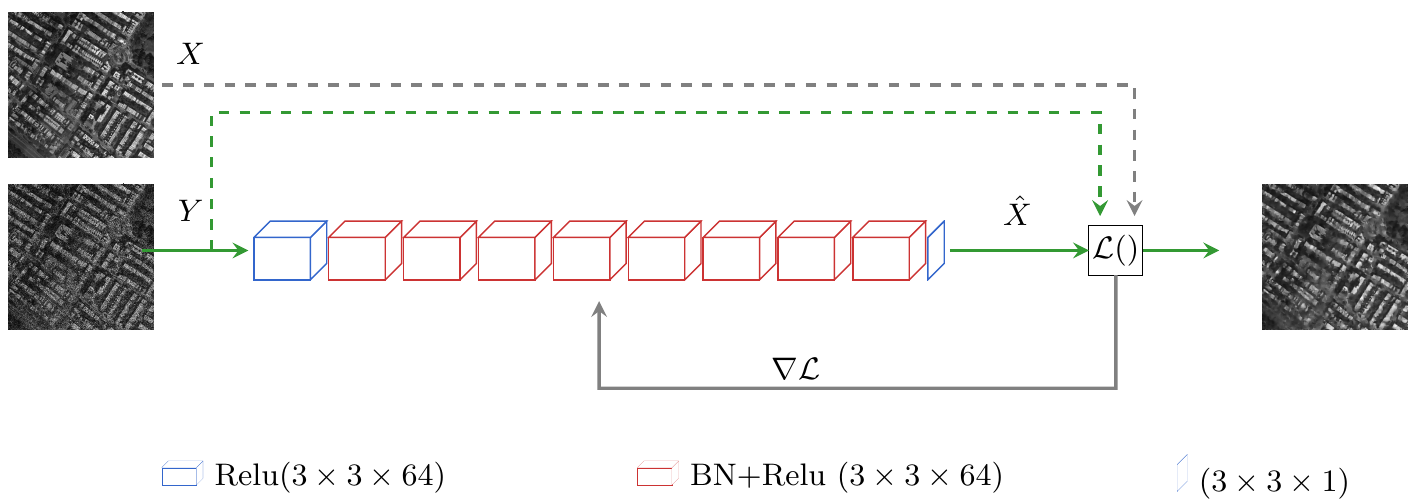}{\figpath net.pdf}
\pgfdeclareimage[width=0.25\textwidth]{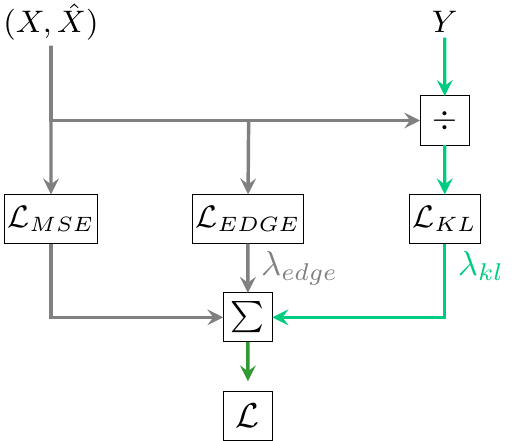}{\figpath cost.pdf}

\begin{document}

\title{ Edge Preserving CNN SAR Despeckling Algorithm}

\author{ S. Vitale\textsuperscript{1}, G. Ferraioli\textsuperscript{2}, V. Pascazio\textsuperscript{1}}

\address{
	\textsuperscript{1 }Università degli Studi di Napoli Parthenope, Dipartimento di Ingegneria, Napoli, Italy - (sergio.vitale, vito.pascazio)@uniparthenope.it\\
	\textsuperscript{2 }Università degli Studi di Napoli Parthenope, Dipartimento di Scienze e Tecnologie,  Napoli, Italy - giampaolo.ferraioli@uniparthenope.it\\
}


\icwg{}   

\abstract{
SAR despeckling is a key tool for Earth Observation. Interpretation of SAR images are impaired by speckle, a multiplicative noise related to interference of backscattering from the illuminated scene towards the sensor. Reducing the noise is a crucial task for the scene understanding. Based on the results of our previous solution KL-DNN, in this work we define a new cost function for training a convolutional neural network for despeckling. The aim is to control the edge preservation and to better filter man-made structures and urban areas that are very challenging for KL-DNN. The results show a very good improvement on the not homogeneous areas keeping the good results in the homogeneous ones. Result on both simulated and real data are shown in the paper. 
}

\keywords{SAR, deep learning, speckle, cnn, denoising}

\maketitle

\section{Introduction}
\label{sec:intro}
Nowadays, a lot of effort has been spending in the direction of earth observation and thanks to the continuous development of satellite sensors more and more data are available. Given the huge availability of data, together with their day by day updating make the use of remote sensing images a crucial source for earth monitoring.
In the last decades, remote sensing images have been used for many applications such as classification, detection and segmentation. The possibility of having an always updated data it is very important for the monitoring of wide areas like forest, agricultural field and urban areas. Moreover, it is also important for detection of natural and man-made disaster like landslide and fire detection.\\
Several methods have been developed to these aims taking advantage from both optical and SAR sensors.
SAR are active sensors that work day and night, in any meteorological condition. Indeed, SAR images are crucial for monitoring in a fast way our planet. SAR imaging formation is characterized by particular geometrical effects: multiple bouncing, shadowing and layover are all related to the presence of an abject on the scene, to its position, to the position and angle of view of the sensor. 
Moreover, SAR images are also affected by a multiplicative noise called speckle \cite{Argenti2013}. Speckle is related to coherent and incoherent interferences among backscattering: depending on the relationship between the roughness of illuminated object and the transmitted wavelength, the backscatterings are spread in several direction and so, backscattering from different objects will interfere each other. Bright pixels are due to the constructive interference, dark pixels to destructive interference. This make the typical alternation of spikes and dark pixels in the SAR image, that obviously impairs the understanding of the scene. Therefore, despeckling usually is used as a preprocessing for further applications. In the last decades several despeckling methods have been developed. The first proposed despeckling filters were the Local filters, so called because work on the assumption of the pixel similarity in its own neighbourhood. These kind of filters \cite{Argenti2013} usually suffer of smoothness in presence of edges. In fact, pixels that belong to the boundary between two different areas do not have many similar in the neighbourhood. In order to overcome this problem, in the last years Non-Local (NL) filters have been proposed  \cite{Deledalle2014}. These filters largely overcome the local one both in noise suppression and edge preservation. They look for similarity in a wider area: given a central pixel in a patch, similar patches are searched in a wider area and the result is the combination of selected patches. The similarity criteria and the combination's rule make the differentiation among different methods \cite{Deledalle2014}, \cite{mambro2018}, \cite{Hossi2018}. Among these filters there are those that make use of optical data for helping the despeckling process \cite{Vitale2019bis} and others that takes advantage from the ratio image \cite{Ferraioli2019bis} (it is the ratio between the SAR image and the filtered one, representing the predicted speckle). The drawback of these NL filters is that they are time consuming.
In the last years, deep learning base methods are showing impressive results in many application of natural image processing such as classification, segmentation and detection \cite{He2017}. Actually, good results are achieved also in several remote sensing application like land classification and segmentation \cite{mazza2019}, super-resolution \cite{Vitale2019c} and detection \cite{gargiulo2019}. 
Clearly, the deep learning base method for despeckling have been proposed \cite{Wang2017}, \cite{Chierchia2017}, \cite{Vitale2019}. Given the great amount of data and the rapidity of producing results easily match with deep learning solution. In this work we proposed a convolutional neural network (CNN) for despeckling. Based on the results of our previous solution \cite{Vitale2019}, we propose a new cost function in order to better preserve and handle the edges.

\section{Data Simulation and Background}
\label{sec: simulation}
Training a CNN for despeckling is a challenging task because of the lack of a noise free reference.
The proposed solution work with simulated data under the fully developed hypothesis of the noise.
In this work we inherit the architecture of KL-DNN \cite{Vitale2019} and we apply another cost function in order to improve the edge preservation: the aim is to better filter not homogeneous areas, such as man-made structures, where KL-DNN performs poorly.
\subsection{Data Simulation}
We simulated a single look ($L = 1$) speckle $N$ under the fully developed hypothesis with the following known Gamma distribution \cite{Argenti2013} 

$$p(N,L) = \frac{1}{\Gamma(L)}L^LN^{L-1}e^{-NL}$$ 

It means, in the simulation we consider just the speckle that homogeneous areas are characterized with.
In order to obtain noise-free reference $X$, we collected images from the optical dataset \textit{Merced Land Use}  \cite{MercedLandUse}, so converted those images to gray scale. So, we multiplied the simulated noise for producing the simulated SAR image $Y = N \cdot X$.
Finally, we tiled the dataset in patches of dimension $64 \times 64$: $30000$ patches were used for the training and $7000$ for the validation.

\subsection{KL-DNN}
\begin{figure}[h]
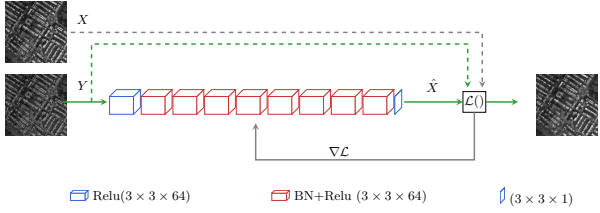

	\centering
	\image{net}
	\caption{KL-DNN architecture}
	\label{fig: net}
\end{figure}

In our previous work, we trained a ten layers CNN (KL-DNN) (for training details refer to \cite{Vitale2019}) on the simulated data. The cost function involved in this work is given by combination of two terms:
$$ \mathcal{L} = \mathcal{L}_{MSE} + \lambda \mathcal{L}_{KL}$$
$$ \mathcal{L}_{MSE} = || \hat{X} - X ||^2$$
$$ \mathcal{L}_{KL} = D_{KL} \left( \frac{Y}{\hat{X}}, \frac{Y}{X}  \right) = D_{KL} (  \hat N, N   )  $$
where $\hat N$ and $N$ are the estimated noise and the theoretical one, respectively;  $D_{KL}(p,q)$ is the Kullback-Leibler (KL) divergence between two distributions $p$ and $q$
$$ D_{KL}(p,q) = \sum_{i} p(i)\log_2\frac{p(i)}{q(i)}$$
In this cost function
$ \mathcal{L}_{MSE} $ is responsible of spatial preservation by comparison of filtered image $\hat{X}$ and the noise-free reference. 
$ \mathcal{L}_{KL} $ is responsible of statistical noise preservation comparing the probabilistic distribution function of estimated $\hat{N}$ and the theoretical one by mean of KL divergence.
The aim of this cost function is to suppress the noise taking care of its statistical properties.
\section{Proposed Method}
The mentioned solution provides good results on homogeneous areas, but presents artefacts in the not homogeneous ones such as urban areas where many man-made structures are present.
Man made structures in real SAR images look totally different from the one in simulated data due to the geometry of SAR image acquisition: when an object is illuminated by the SAR, effects like layover, shadowing and multiple bounces arise. Usually, in SAR image a building is characterized by a side with a strong backscattering due to multiple reflections with the ground, and the other side is darker due to the layover and shadowing. 
The speckle in such areas is not fully developed \cite{Frery97}, and our simulated data do not include such effects and statistics. 

Given KL-DNN works under the fully developed hypothesis, it does not know how to filter man-made structures. Generally, it is going to filter them in order to produce a speckle that is fully developed and so many artefacts arise.

In order to limit this problem we include a term in the cost function for improving the edge preservation. The aim is to make the network able to recognize man-made structures and to preserve their shape (usually characterized by strong edges).
In such way, we want the network to filter homogeneous areas and to preserve objects details.
So the actual cost function is composed of three terms:

\begin{figure}[h]
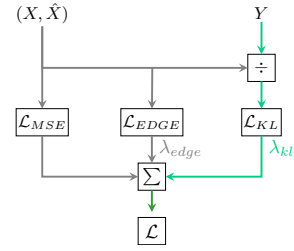

	\centering
	\image{cost}
	\caption{Proposed Cost Function}
	\label{fig:cost function}
\end{figure}

$$ \mathcal{L} = \mathcal{L}_{MSE} + \lambda_{kl} \mathcal{L}_{KL} + \lambda_{edge} \mathcal{L}_{EDGE} $$

where $$\mathcal{L}_{EDGE} = \left( \de{X}{u} - \de{\hat{X}}{u} \right) ^2 +  \left( \de{X}{v} - \de{\hat{X}}{v} \right) ^2  $$

where $\de{I}{u}$ and $\de{I}{v}$ are respectively the derivatives along the rows and columns of the image $I$.
With this function we train the network to suppress the noise, taking care both of the statistical properties of the speckle and of the present edges.

\section{Experiments}
In order to have a fair comparison with KL-DNN, we  use its same architecture. Moreover, we trained both the proposed network and KL-DNN on the same dataset, with Adam optimizer \cite{Kingma14}.

Numerical and visual assessment are carried out for validating the method. We test our solution on both simulated and real data. We show comparison with our previous solution KL-DNN in order to show the impact of the cost function. Moreover, for sake of completeness we also compare with two famous non-local filters such as FANS \cite{Cozzolino2014} and SAR-BM3D \cite{Parrilli2012}. 
The simulated data are taken from the Mercedes dataset and from scraped Google Maps \cite{Wang2017} and never seen during the training. The simulation follows the process depicted in Section \ref{sec: simulation}. 

\begin{figure}[h]
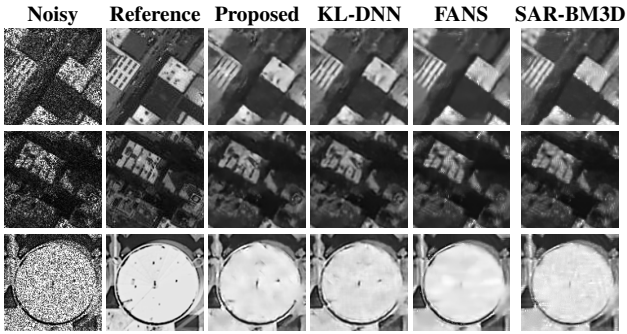

	\centering
	\begin{tabular}{cccccc}
		\labNoise &  \labRef &  \labProp & \labKldnn &\labFans  & \labSarbm\\
		\image{sim1} & \image{ref1} & \image{prop1} & \image{kl1} & \image{fans1}&\image{sarbm3d1}\\
		\image{sim2} & \image{ref2} & \image{prop2} & \image{kl2} & \image{fans2}&\image{sarbm3d2}\\
		\image{sim3} & \image{ref3} & \image{prop3} & \image{kl3} & \image{fans3}&\image{sarbm3d3}\\
		
	\end{tabular}
	\caption{Results on simulated data, from top to bottom: clip1 (scraped Google Maps), clip2 (scraped Google Maps), clip3 (Mercedes Land Use)}
	\label{fig: results_sim}
\end{figure}

In Fig.\ref{fig: results_sim} are shown results on simulated data. In all cases, it can be appreciated how the introduction of the term $\mathcal{L}_{EDGE}$ in the cost function improves the edge preservation. In clip1 and clip2 (first two rows of Fig.\ref{fig: results_sim}), we consider an urban area: compared to KL-DNN, the filtered image is closer to the reference: the results are sharper and more clean. Moreover, the proposed approach shows a much better edge and detail preservation, e.g. the objects on the rooftop are more visible than in KL-DNN.  Same consideration can be done on the image for clip3 (last row of Fig.\ref{fig: results_sim}) that shows a storage tanks: compared to KL-DNN, the edges are better preserved and details look sharper. In both cases, FANS and SAR-BM3D show several artefacts: FANS tends to preserve edges but is oversmoothed and lose a lot of details., SAR-BM3D better preserves details with respect to FANS but smooths the edges. Anyway, the proposed solution shows better edges and details preservation with respect the other methods.

\begin{table}[h]
	\setlength{\tabcolsep}{5mm}
	\centering
	\begin{tabular}{c|ccc}
		Clip1        & SSIM & SNR & MSE \\
		\hline
		\labFans  &  0.744   & 8.59  & 287 \\
		\labSarbm &  0.763   & 8.67  & 282 \\
		\hline
		\labKldnn &  0.767   & 8.79  & 274\\
		\labProp  &  \aa{0.769}  &  \aa{8.83}  & \aa{272}\\
		\hline
	\end{tabular}
	\caption{Numerical Results on clip1}
	\label{tab:clip1}
\end{table}

\begin{table}[h]
	\centering
	\setlength{\tabcolsep}{5mm}
	
	\begin{tabular}{c|ccc}
		Clip2        & SSIM & SNR & MSE \\
		\hline
		\labFans  &  0.785   & 7.93  & 265 \\
		\labSarbm &  0.793   & 7.86  & 270 \\
		\hline
		\labKldnn &  0.794   & 7.97  & 263\\
		\labProp  &  \aa{0.796}  &  \aa{7.98}  & \aa{262}\\
		\hline
	\end{tabular}
	\caption{Numerical Results on clip2}
	\label{tab:clip2}
\end{table}

\begin{table}[h]
	\centering
	\setlength{\tabcolsep}{5mm}
	
	\begin{tabular}{c|ccc}
		Clip3        & SSIM & SNR & MSE \\
		\hline
		\labFans  &  0.758   & 9.67  & 419 \\
		\labSarbm &  0.729   & 8.72  & 522 \\
		\hline
		\labKldnn &  0.761   & 9.89  & 398\\
		\labProp  &  \aa{0.796}  &  \aa{10.04}  & \aa{385}\\
		\hline
	\end{tabular}
	\caption{Numerical Results on clip3}
	\label{tab:clip3}
\end{table}

These considerations are confirmed by the numerical assessment in Tabb. \ref{tab:clip1}-\ref{tab:clip3}.
The presented metrics indicate how much the filtered image is close the reference one (MSE), how much the noise is suppressed (SNR) and how much the filtered and reference image are similar from a perceptual point of view (SSIM). Ideal filter will give MSE=0, SNR=$\inf$ and SSIM=1.
In all the metrics, the proposed solution outperforms the other methods validating the previous consideration.

\begin{figure}[h]
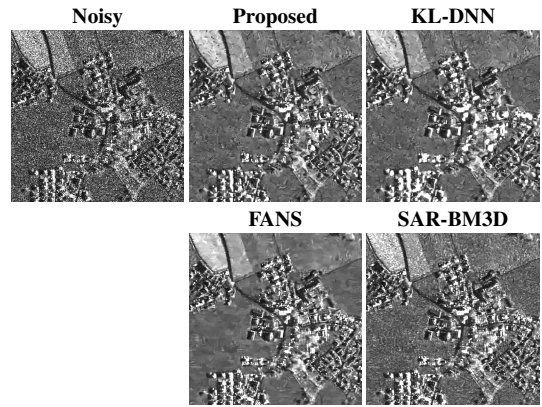

	\centering
	\begin{tabular}{ccc}
		\labNoise &    \labProp & \labKldnn\\
		\image{ros_noisy} & \image{ros_prop} &  \image{ros_kldnn}\\
		 				  &\labFans  & \labSarbm \\
		  				& \image{ros_fans}&\image{ros_sarbm3d}\\
	\end{tabular}
	\caption{Results on Real Data: Rosenheim area taken from Terrasar-X}
	\label{fig: results_real}
\end{figure}

\begin{figure}[h]
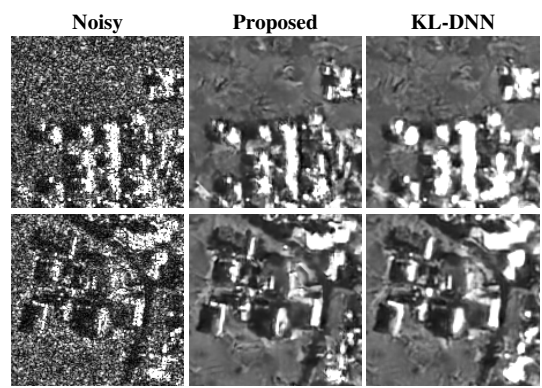

	\centering
	\begin{tabular}{ccc}
		\labNoise &    \labProp & \labKldnn \\
		\image{ros_noisy1} & \image{ros_prop1} &  \image{ros_kldnn1}\\
		\image{ros_noisy2} & \image{ros_prop2} &  \image{ros_kldnn2}\\
		
	\end{tabular}
	\caption{Details of real data}
	\label{fig: details_real}
\end{figure}

Regarding the real SAR images we consider a TerraSAR-X image taken from Rosenheim . In Fig. \ref{fig: results_real} are shown the results for all the methods. Generally, we can keep the considerations done for simulated data. In this case, the proposed solution is sharper than KL-DNN showing a better edge preservation. Moreover, the proposed solution better preserves details and small object that are completely lost in FANS. Instead SAR-BM3D has very good edge preservation but a poor noise reduction.
Actually, we want to focus on those challenging areas for KL-DNN. It means we want to find out the behaviour on man-made structures where the speckle is not fully developed and where KL-DNN tends to smooth the image.
In Fig. \ref{fig: details_real} two details from Rosenheim are shown. In both cases it can be noted how KL-DNN face difficulties in filtering such areas and tends to smooth them. After all, KL-DNN is trained under the fully developed hypothesis, so these troubles should be expected . Watching the proposed results it can be noted that, introducing a cost function for edge preservation helps the network in localize and recognize these strong backscatterers as object to be preserved and so the smoothing effect is strongly preserved. It is clear from the two details in Fig.\ref{fig: details_real} and the whole image in Fig.\ref{fig: results_real} that the proposed solution is able to better preserve the sharpness of these edges without losing details on the homogeneous areas, even if the assumption of fully developed hypothesis is still valid during the training.  
So, the term $\mathcal{L}_{EDGE}$ helps the network in overcoming the limitation of fully developed hypothesis by preserving the geometry of man-made structures.

\section{Conclusions}	
In this paper a convolutional neural network for despeckling has been proposed. In this work we define a new cost function, based on the knowledge of our previous solution KL-DNN where the network is trained under the fully developed hypothesis. This cost function aims to better preserve the edges and to have a better filtering process in areas where man-made structures are present. The results show how, even if the network is still trained under the fully developed assumption, the introduction of a loss taking care of the edges helps the filter in treating the not homogeneous areas.

{
	\begin{spacing}{1.17}
		\normalsize
		\bibliography{ISPRSguidelines_authors} 
	\end{spacing}
}

%
%

\end{document}